\documentclass{itg}
\usepackage{amsmath}
\usepackage{amssymb}
\usepackage[dvipdfm]{hyperref}

\input{itg06_notations}

\begin{document}
\title{Precoding for $2\times2$ Doubly--Dispersive WSSUS Channels}
\author{Peter Jung\\
  Fraunhofer German-Sino Lab for Mobile Communications (MCI) and the Heinrich-Hertz Institute \\[.1em]
  jung@hhi.fraunhofer.de}
\abstract{Optimal link adaption to the scattering function of wide sense stationary uncorrelated scattering (WSSUS)
  mobile communication channels is still an unsolved problem despite its importance 
  for next-generation system design. In multicarrier transmission such link adaption is performed by pulse
  shaping which in turn is equivalent to precoding with respect to the second order channel statistics. 
  In the present framework a translation of the precoder optimization problem
  into an optimization problem over trace class operators is used \cite{jung:isit05,jung:wssuspulseshaping}. 
  This problem which is also well-known in the
  context of quantum information theory is unsolved in general due to its non-convex nature. 
  However in very low dimensions the problem formulation reveals an additional analytic structure 
  which admits the solution to the optimal precoder and multiplexing scheme.
  Hence, in this contribution  the analytic solution of the problem for  
  the $2\times2$ doubly--dispersive WSSUS channel is presented.}
\maketitle
\section{Introduction}
It is well known, that channel information at the transmitter increases link capacity. 
However, since future mobile communication is expected to operate in fast varying channels,
it is not realistic to assume perfect channel knowledge at the transmitter. On the other hand statistical 
information can be used which does not change in the rapid manner as the channel itself.
In multicarrier communications this can be employed for the design of transmitter
and receiver pulse shapes. However, the problem of optimal signaling in this context is
still an unsolved problem.

In this paper the most basic case of precoder and equalizer optimization with respect
to the statistics of a doubly--dispersive channel is considered. Hence, the focus is
on a model in which two random complex symbols (iid and zero mean distributed) 
are transmitted parallel over a $2\times2$ random channel. 
Before transmission a special kind of linear precoding is performed which is motivated
from Weyl--Heisenberg signaling scheme.
The channel itself is
a randomly weighted superposition of four possible channel operations, which are
\begin{enumerate}
\item do not change anything 
\item permutation in the time domain
\item permutation in the frequency domain
\item permutation in the time and frequency domain
\end{enumerate}
Each action is distributed independently from the others and  scaled
with a certain amount of power. What is
the optimal precoding and multiplexing scheme? It will be shown that this is the formulation of
the WSSUS pulse shaping problem for Weyl--Heisenberg signaling in the lowest possible dimension.

The paper is organized as follows. In the first part Weyl--Heisenberg (or Gabor) signaling in 
$\Ltwo(\mathbb{R})$ and $\mathbb{C}^L$ is introduced.
Then the main optimization functional according to \cite{jung:isit05,jung:wssuspulseshaping}
is established. It will open up the relation to the pure state channel fidelity optimization which 
is an ongoing research topic in quantum information theory. In the main part 
the problem is solved for $\mathbb{C}^2$.

\section{Signal Model}
\subsection{Weyl--Heisenberg Signaling on $\Ltwo(\mathbb{R})$}

In this article a transmit baseband signal $s(t)$ is considered which is
a superposition of time--frequency translates of a single 
prototype function $\gamma(t)$ with $\lVert\gamma\rVert_2=1$. The translations are according to a subset 
of a lattice $\Lambda\Fieldtwo$ in the time--frequency plane, i.e.
\begin{equation}
   \begin{aligned}
      s(t)
      &=\sum_{n\in\Indexset}x_n(\Shift_{\Lambda n}\,\gamma)(t)
   \end{aligned}
   \label{equ:txsignal}
\end{equation}
where in this constellation $\Field=\mathbb{Z}$ and $\Lambda$ denotes 
the $2\times2$ generator matrix. The indices $n=(n_1,n_2)$  range over 
the doubly-countable set $\Indexset\subset\Fieldtwo$, referring to the data burst to be transmitted. 
Let
\begin{equation}
   \begin{aligned}
      (\Shift_\mu f)(t)\defeq e^{i2\pi\mu_2 t}\gamma(t-\mu_1)
   \end{aligned}
\end{equation}
denote the time-frequency shift operator (or phase space displacement operator).
It is well-known that the operators $\Shift_\mu$ establish 
an unitary representation of the polarized Heisenberg group and up to 
phase factors they are equal to the Weyl operators (see \cite{folland:harmonics:phasespace}). 
The complex data symbols to transmit are $x_n$ where $n_1$ is the time instant
and $n_2$ is the subcarrier index.
The transmit signal is passed through a linear time-variant channel denoted by the operator $\BH$  
and further distorted by an
additive white Gaussian noise process $n(t)$.
Hence, the received signal is then
\begin{equation}
   \begin{aligned}
      r(t)=(\BH s)(t)+n(t)
      =\int \BHSpread(\mu)(\Shift_{\mu}s)(t)d\mu + n(t)
   \end{aligned}
   \label{eq:rxsignal}
\end{equation}
with $\BHSpread(\mu)$ being a realization of the channel {\it spreading
  function}. In practice  $\BHSpread(\mu)$ is causal and has finite support.
The second order statistics of $\BHSpread(\mu)$ is
\begin{equation}
   \EX{\BHSpread(\mu)
     \overline{\BHSpread(\nu)}}=\BHScat(\mu)\delta(\mu-\nu)
\end{equation}
where $\BHScat(\mu)$ is the {\it  scattering function} with
$\lVert\BHScat\rVert_1=1$ (no overall path-loss).
To obtain the data symbol $\tilde{x}_m$ the receiver does the
projection 
\begin{equation}
   \tilde{x}_m=\langle \Shift_{\Lambda m}g,r\rangle=\int\overline{(\Shift_{\Lambda m}g)(t)}r(t)dt
\end{equation}
onto a time-frequency shifted version of the function $g(t)$ (i.e. an equalization filter) with $\lVert g\rVert_2=1$. 
Let
\begin{equation}
   \begin{split}
      H_{m,n}
      &\defeq\langle \Shift_{\Lambda m}g,\BH\Shift_{\Lambda n}\gamma\rangle
   \end{split}
\end{equation}
be the elements of the channel matrix $H\in\mathbb{C}^{\Indexset\times\Indexset}$ and the noise
$n_m\defeq\langle \Shift_{\Lambda m}g,n\rangle$
the transmission scheme can be formulated as the linear equation $\tilde{\Bx}=H\Bx+\Bn$.

\subsection{Weyl--Heisenberg Signaling on $\mathbb{C}^L$}
With the connection of shift operators to unitary representations of
the Weyl-Heisenberg group
it is straightforward to pass over to finite dimensional models, i.e.
(cyclic) shift operators on $\mathbb{C}^L$ which are given as unitary representation
of finite Heisenberg groups. Hence, let $\MatrixAlg{L}{\mathbb{C}}$ 
be the algebra of $L\times L$ matrices over $\mathbb{C}$. Then 
the operators $\Shift_\mu\in\MatrixAlg{L}{\mathbb{C}}$ are given as the matrices
\begin{equation}
   \left(\Shift_{\mu}\right)_{mn} =\delta_{m,n+\mu_1}e^{i\frac{2\pi}{L}\mu_2(m-1)}
\end{equation}
where all index--arithmetics are modulo $L$. The finite Heisenberg group is then
\begin{equation}
   \HGgroup_L=\{\Shift_{(m,n)}| m,n\in\Field_L\}
\end{equation}
where $\Field_L=\{0\dots L-1\}$.


\section{Formulation of the  Problem}
In the view of multicarrier transmission only single carrier equalization is considered.
Interference cancellation is not used in this field due to complexity reasons.
Hence it is then naturally to require $a$ (the channel gain of the lattice point
$m\in\Indexset$) to be maximal 
and the interference power 
$b$ from all other lattice points to be minimal as possible, where
\begin{equation}
   a\defeq|H_{m,m}|^2\,\,\,\,\text{and}\,\,\,\,
   b\defeq\sum_{n\neq m}|H_{m,n}|^2  
\end{equation}
This addresses the concept of 
{\it pulse shaping}, hence to find jointly good pulses $\{g,\gamma\}$ 
(or precoders and equalizers) achieving
maximum channel gain and minimum interference power. 
A comprehensive framework for the optimization of redundant precoders and equalizers with respect to
instantaneous time-invariant channel realizations (assumed to be known at the transmitter) 
is given in \cite{scaglione:redfilt:partA}.
However in certain scenarios it is much more realistic 
to adapt the pulses only to the second order statistics, given by $\BHScat(\mu)$ 
and not to a particular realization $\BHSpread(\mu)$.
This is in the sense of defining for the considered time-frequency slot $m$
\begin{equation}
   \SINR(g,\gamma,\Lambda)\defeq\frac{\Ex{\BH}{a}}{\sigmaN+\Ex{\BH}{b}}
   \geq\frac{\Ex{\BH}{a}}{\sigmaN+B_\gamma-\Ex{\BH}{a}}
   \label{eq:sinrdef}
\end{equation}
as a {\it long term} performance measure. 
Optimal signaling via (\ref{equ:txsignal}) and (\ref{eq:rxsignal}) maximizing (\ref{eq:sinrdef}) 
\emph{independent} of $\BHSpread(\cdot)$
is of central relevance for band efficient and low-complexity multicarrier implementation.
For example, results on joint multipath and Doppler diversity critically rely on so called
approximate $\BHSpread$-independent basis expansions proposed in
\cite{sayeed:jointdiversity} of $\BH$. But a general approach how to obtain the 
''best basis'' for $\BH$ being WSSUS operators, i.e. maximizing (\ref{eq:sinrdef}),  is still unknown.
Nevertheless, some iterative methods are contained in \cite{jung:spawc2004}.

The derivation of the lower bound in (\ref{eq:sinrdef}) can be found in \cite{jung:ieeecom:timevariant}
and in the context of pulse shaping in \cite{jung:wssuspulseshaping}. Equality is
achieved if the set $\Gabor(\gamma,\Lambda,\Fieldtwo)\defeq\{\Shift_{\Lambda n}\gamma\}_{n\in\Fieldtwo}$ 
(called a Gabor set or family) establishes a tight (Gabor or Weyl--Heisenberg) 
frame \cite{feichtinger:gaborbook,jung:wssuspulseshaping}. The constant $B_\gamma$ is
called the Bessel bound of $\Gabor(\gamma,\Lambda,\Fieldtwo)$ and related to its redundancy.
For $\Gabor(\gamma,\Lambda,\Fieldtwo)$ being an ONB it follows $B_\gamma=1$.

Straightforward computation yields the {\it channel fidelity} (or averaged gain term) given as
\begin{equation}
   \Ex{\BH}{a}=\int\BHScat(\mu)|\langle g,\Shift_{\mu}\gamma\rangle|^2d\mu
\end{equation}
and the averaged interference power
\begin{equation}
   \Ex{\BH}{b}=\sum_{m\neq 0}\int\BHScat(\mu)|\langle g,\Shift_{\Lambda m+\mu}\gamma\rangle|^2d\mu
\end{equation}
Hence, the $\SINR(g,\gamma,\Lambda)$ is independent of $m$. Let $\Shift_\mu^*$ denote
the hermitian adjoint operator of $\Shift_\mu$ with respect to $\langle\cdot,\cdot\rangle$. Then the
channel fidelity can be rewritten as
\begin{equation}
   \begin{aligned}
      \Ex{\BH}{a}
      &=\langle g, 
      \left[\int\BHScat(\mu)\Shift_{\mu}\Gamma\Shift^*_{\mu}d\mu\right]
      g\rangle
      \defeq\Trace{A(\Gamma)G}
   \end{aligned}
\end{equation}
where $G$ (and $\Gamma$) is the (rank-one) orthogonal projector onto $g$ (and $\gamma$), i.e.
$Gf\defeq \langle g,f\rangle g$. 
Similarly 
\begin{equation}
   \begin{aligned}
      \Ex{\BH}{b}
      &=\Trace{ 
        \left[
          \sum_{m\neq 0}\Shift_{\Lambda m}
          A(\Gamma)
          \Shift^*_{\Lambda m}
        \right]G}
      \defeq\Trace{C(\Gamma)G}
   \end{aligned}
\end{equation}
where $A(\cdot)$ and $C(\cdot)$ are affine maps
acting on linear operators. 
The definition of $A(\cdot)$ in particular is also known as Kraus representation
of a {\it completely positive map} (see for example
\cite{choi:cpmaps}) which establishes a relation to 
quantum channels. Due to $\lVert\BHScat\rVert_1=1$ and $\Shift_{\mu}$ being unitary operators, 
the following properties can be verified
\begin{equation}
   \begin{aligned}
      A\,\text{is unital}\,&\Leftrightarrow A(\Id)=\Id \\
      A\,\text{is trace preserving}\,&\Leftrightarrow \Trace{ A(X)}=\Trace{X} \\
      A\,\text{is hermiticity preserving}\,&\Leftrightarrow A(X^*)=A(X)^*\\
      A\,\text{is entropy increasing}\,&\Leftrightarrow A(X)\minor X
   \end{aligned}
\end{equation}
where $\minor$ is in the finite case the partial order due to eigenvalue majorization
Thus, $A(\cdot)$ flatten the eigenvalue distribution of its input (increasing its entropy).
After application of $A(\cdot)$ onto a rank-one projector $\Gamma$, the ''output in the averaged
sense'' (over an ensemble of WSSUS channels) $A(\Gamma)\minor\Gamma$ is not (in general) rank-one. In this picture 
so called additional eigen modes occur which can not be collected together using a rank-one 
equalizer (a single equalization filter). 

With $D(\Gamma)\defeq C(\Gamma+\sigmaN\Id)$  the $\SINR$ optimization problem reads
\begin{equation}
   \begin{aligned}
      \max_{G,\Gamma\in\setrankone,\Lambda}{\SINR(G,\Gamma)}
      =\max_{G,\Gamma\in\setrankone,\Lambda}{\frac{\Trace{A(\Gamma)G}}{\Trace{D(\Gamma)G}}}
   \end{aligned}
   \label{eq:sinrdef:trace}
\end{equation}
The maximization is performed over possible lattices $\Lambda$ and $G,\Gamma\in\setrankone$,
where $Z$ denotes the set of orthogonal rank-one projectors, i.e.
\begin{equation*}
   \begin{aligned}
      \settraceideal&\defeq\{z\,|\,\Trace\,z=1,z^*=z,z\geq 0\} \\
      \setrankone   &\defeq\{z\,|\,z\in\settraceideal,z^2=z\}
   \end{aligned}
\end{equation*}
Note that the convex hull of $\setrankone$ is the subset $\settraceideal$ of positive--semidefinite 
trace class operators.

The maximizing $G$ for fixed $\Gamma$ in (\ref{eq:sinrdef:trace}) is achieved by an
orthogonal projection onto the generalized eigenspace
corresponding to the maximal generalized eigenvalue 
$\lambda_{\max}(A(\Gamma),D(\Gamma))$. Thus it remains the ''transmitter--side only'' optimization:
\begin{equation}
   \begin{aligned}
      \max_{\Gamma\in\setrankone,\Lambda}{\SINR(\Gamma)}=
      \max_{\Gamma\in\setrankone,\Lambda}\lambda_{\max}(A(\Gamma),D(\Gamma))
   \end{aligned}
   \label{eq:sinr:max:rankone}
\end{equation}
With the definition of an adjoint channel it also possible to obtain a ''receiver--side only'' 
optimization \cite{jung:wssuspulseshaping}.
Due to joint quasi-convexity of the function $\lambda_{\max}(\cdot,\cdot)$ (see for example \cite{boyd:convexopt}) 
the constraint $\Gamma\in\setrankone$ can be relaxed to the convex set $\settraceideal$. 
Thus, the optimization problem is identified as {\it convex constrained quasi-convex maximization}.
Furthermore, if the inverse of $D(\Gamma)$ exists, the problem can be rewritten as
a classical eigenvalue problem. Note that this is non-convex optimization.

Now, the lower bound in (\ref{eq:sinrdef}) suggests the maximization of the 
channel fidelity $\Trace{A(\Gamma)G}$ only, i.e. 
\begin{equation}
   \begin{aligned}
      \max_{G,\Gamma\in\setrankone}{\Trace{A(\Gamma)G}}=\max_{\Gamma\in\settraceideal}{\lambda_{\max}(A(\Gamma))}\leq1
   \end{aligned}
   \label{eq:gainmax}
\end{equation}
which does not depend on the lattice $\Lambda$.
The derivation from the left to the right side in (\ref{eq:gainmax})
is again due to convexity of $\lambda_{\max}(\cdot)$, linearity of $A(\cdot)$ and unitarity of $\Shift_\mu$.
It can be shown that this formulation is now equivalent to the problem of maximizing the 
{\it quantum channel fidelity} \cite{jozsa:fidelity} for $G$ being a {\it pure state} (rank-one).
Where the solution of (\ref{eq:gainmax}) for single--dispersive channels is straightforward,
the general case of this optimization problem -- convex constrained convex maximization -- 
is unsolved in general. For $\BHScat(\mu)$ being a 
two--dimensional Gaussian the solution was found in \cite{arxiv:0409063}. Another proof which 
additional gives the uniqueness of the solution is in \cite{jung:isit06}.

Already in \cite{jung:wssuspulseshaping} it is conjectured that with a proper selection of 
a basis for the $L^2$ dimensional real vector space of hermitian operators on $\mathbb{C}^L$ the left side of  
(\ref{eq:gainmax}) could be rewritten as 
a bilinear program over so called $L^2-1$ dimensional Bloch manifolds $\setblochrankone(L)$ \cite{arxiv:0502153}, 
i.e.
\begin{equation}
   \begin{split}
      \max_{X,Y\in\setrankone}\Trace{A(X)Y}=\max_{x,y\in\setblochrankone(L)}\langle x,ay\rangle
   \end{split}
\end{equation}
where $(a_{ij})\in\MatrixAlg{L^2}{\mathbb{R}}$ is then the matrix representation of $A(\cdot)$ in this basis.
Indeed -- this parameterization will be used in the next section.

Finally, if the Gabor set $\Gabor(\Gamma,\Lambda,\Fieldtwo)$ with the ''channel fidelity''--optimal $\Gamma$ 
would establish a tight frame for some lattice $\Lambda$, the solution maximizes $\SINR(\Gamma)$ too.
In this case ''channel fidelity''--maximization equals minimization of the averaged interference.
Normally this is not the case and it remains lattice optimization with \mbox{$\det\Lambda=\text{const}$}. 
In pulse shaping procedures then
a so called orthogonalization with respect to $\Lambda$ has to be applied on $\Gamma$
to minimize the Bessel bound $B_\gamma\geq 1$  \cite{strohmer:lofdm2,jung:wssuspulseshaping}.
But it will turn out that this step is not needed here  for the $\mathbb{C}^2$ case (at $\det\Lambda=1$).

\section{The \mbox{$2\times2$} WSSUS Channel}
For this simple toy model $\mathbb{C}^2$ as the underlying Hilbert space is assumed, i.e. \mbox{$L=2$}.
The corresponding finite Heisenberg group is
\mbox{$\HGgroup_2=\{\Shift_{(0,0)},\Shift_{(1,0)},\Shift_{(0,1)},\Shift_{(1,1)}\}$}, where
\begin{equation}
   \begin{split}
      \Shift_{(0,0)}&=\left(\begin{array}{cc}
           1 &  0 \\
           0 &  1 \\
        \end{array}\right)
      \Shift_{(0,1)}=\left(\begin{array}{cc}
           1 &  0 \\
           0 &  -1 \\
        \end{array}\right)\\
      \Shift_{(1,0)}&=\left(\begin{array}{cc}
           0 &  1 \\
           1 &  0 \\
        \end{array}\right)
      \Shift_{(1,1)}=\left(\begin{array}{cc}
           0 &  1 \\
           -1 &  0 \\
        \end{array}\right)\\
   \end{split}
\end{equation}
As already intended in the introduction these matrices represent four basic channel operations
which occur in a randomly weighted superposition. The matrix  $\Shift_{(1,0)}$ represents the only
cyclic shift that exists, hence it switches the input samples. The element $\Shift_{(0,1)}$ does
the same in frequency domain. With $\Shift_{(1,1)}$ both operations occur simultaneously 
and $\Shift_{(0,0)}$ does not change the input at all.
The following relations are important
\begin{equation}
   \begin{split}
      \Shift_{(0,0)}=\sigma_0 \quad& 
      \Shift_{(0,1)}=\sigma_3=F\Shift_{(1,0)}F^*\\
      \Shift_{(1,0)}=\sigma_1\quad& 
      \Shift_{(1,1)}=i\sigma_2=\Shift_{(0,1)}\Shift_{(1,0)}\\
   \end{split}
   \label{eq:shift:pauli}
\end{equation}
where the $\sigma_i$ are the well known Pauli-matrices and $F$ is the 2x2 Fourier matrix given as 
\begin{equation}
   \begin{split}
      F\defeq\frac{1}{\sqrt{2}}\left(\begin{array}{cc}
           1 &  1 \\
           1 &  -1 \\
        \end{array}\right)
   \end{split}
\end{equation}
Furthermore we define the following constants
\begin{equation}
   \begin{split}
      p_0\defeq \BHScat(0,0) & \qquad p_1\defeq\BHScat(1,0)\\ 
      p_2\defeq \BHScat(1,1) & \qquad p_3\defeq\BHScat(0,1)
   \end{split}
\end{equation}
given by the four possible values of scattering function $\BHScat(\mu)$. 
This allows us to write 
\begin{equation}
   \begin{split}
      A(X)=\int \Shift_\mu X\Shift_\mu^* \BHScat(\mu)d\mu=\sum_{i=0}^3 p_i\sigma_i X\sigma_i^*
   \end{split}
\end{equation}
The Pauli-matrices establish an orthogonal basis for the real vector space of hermitian 2x2 matrices
with inner product $\langle X,Y\rangle=\Trace{X^*Y}$. Thus,
every 2x2 hermitian matrix $X$ has a decomposition $X=\frac{1}{2}\sum_{i=0}^3x_i\sigma_i$ with $x_i\in\mathbb{R}$.
Furthermore they establish up to factors the finite Weyl--Heisenberg group itself as shown in (\ref{eq:shift:pauli}).
This additional property will admit the direct solution of the problem.
The following properties are useful to verify the calculations later on: 

\begin{equation}
   \begin{aligned}
      \sigma_i^2&=\sigma_0\\
      \Trace{\sigma_i}&=2\delta_{i0} \\
      \det{\sigma_i}&=-1\quad\text{for}\quad i=1,2,3\\
      \sigma_i\sigma_j &= \begin{cases}
         \sigma_i & j=0\\
         \sigma_j & i=0\\
         i\epsilon_{ijk}\sigma_k + \delta_{ij}\sigma_0 & i,j\neq0 
      \end{cases}\\
      \Trace{\sigma_i\sigma_j}&=2\delta_{ij}
   \end{aligned}
\end{equation}
where $\epsilon_{ijk}$ is the Levi-Civita symbol.
Note that the Pauli-matrices are unitary \emph{and} hermitian.

\subsection{The Channel Fidelity}
Recall the equivalent problem formulation
\begin{equation}
   \begin{split}
      \max_{X,Y\in\setrankone}\Trace{A(X)Y}=\max_{x,y\in\setblochrankone(2)}\langle x,ay\rangle
   \end{split}
   \label{eq:twobytwoproblem}
\end{equation}
where $\setblochrankone(2)=\{x|\frac{1}{2}\sum_{i}x_i\sigma_i\in\setrankone\}$ is a $3$--dimensional 
sub-manifold in 
$\mathbb{R}^4$ --- the Bloch manifold for 2x2. The matrix $(a_{ij})\in\MatrixAlg{4}{\mathbb{R}}$ is the corresponding matrix 
representation of $A(\cdot)$ with elements 
\begin{equation}
   a_{ij}=\frac{1}{4}\Trace{A(\sigma_i)\sigma_j}
\end{equation}
The Bloch parameterization is a well known tool in quantum physics which admits for $L=2$ the
simple interpretation of $3$-dimensional Bloch vectors.
Because it is not very common in this context a short overview will be given.
First let us evaluate the conditions for a vector $x\in\mathbb{R}^4$ to be in 
$\setblochrankone(2)$. We will adopt the notation $\vec{\cdot}$ which means
$\vec{x}=(x_1,x_2,x_3)$ for $x=(x_0,x_1,x_2,x_3)\in\mathbb{R}^4$.
\begin{mylemma}
   A real vector $x\in\mathbb{R}^4$ is in $\setblochrankone(2)$ iff
   $x_0=1$ and $\lVert\vec{x}\lVert_2=1$.
   \label{lemma:twobytwo:blochmanifold}
\end{mylemma}
\vspace*{.3em}
\noindent Thus, it is the fact that the Bloch manifold 
\begin{equation*}
   \setblochrankone(2)=\{x=(x_0,\vec{x})| x_0=1\quad\text{and}\quad \lVert\vec{x}\lVert_2=1\}
\end{equation*}
is a 2-sphere in $\mathbb{R}^4$ with origin $(1,0,0,0)$. To visualize why it is not easy 
to extent this concept to higher dimensions we give a short proof.\\[.5em]
\begin{proof}
   We have to proof $\Trace{X}=\Trace{X^2}=1$ and $X\geq0$. 

   The trace normalization represents the $l_2$ normalization of the precoders and equalizers.
   The condition
   \begin{equation}
      \begin{split}
         \Trace{X}
         &=\frac{1}{2}\Trace{\sum_ix_i\sigma_i}
         =\frac{1}{2}\sum_ix_i\Trace{\sigma_i}=
         \frac{1}{2}x_0\cdot 2
      \end{split}         
   \end{equation}
   is fulfilled if $x_0=1$.
   The second trace requirement is the rank--one constraint if $X\geq0$, i.e. the condition
   \begin{equation}
      \Trace{X^2}=\frac{1}{4}\sum_{i,j}\Trace{(x_i\sigma_i)(x_j\sigma_j)}=
      \frac{1}{2}\lVert x\rVert^2_2
   \end{equation}
   is fulfilled if $\lVert x\rVert^2_2=2$.
   Using $\lVert x\rVert_2^2=x_0^2+\lVert\vec{x}\rVert_2^2=1+\lVert\vec{x}\rVert_2^2=2$ gives the requirement:
   \begin{equation}
      \lVert\vec{x}\rVert_2^2=1
   \end{equation}
   To ensure $X\geq0$  we need conditions on the determinants. Firstly 
   \begin{equation}
      \begin{aligned}
         \det X&=\frac{1}{4}\det\sum_i x_i\sigma_i\\
         &=\frac{1}{4}\det{
           \left(\begin{array}{cc}
                x_0+x_3 &  x_1-ix_2 \\
                x_1+ix_2 &  x_0-x_3 \\
             \end{array}\right)}\\
         &=\frac{1}{4}\left(x_0^2-\lVert\vec{x}\rVert_2^2\right)\geq 0
      \end{aligned}
   \end{equation}
   which is automatically fulfilled due to $x_0=1$ and $\lVert\vec{x}\rVert=1$.
   Secondly the upper left sub--determinant is
   \begin{equation}
      \frac{1}{4}\det(x_0+x_3)=\frac{1}{4}(x_0+x_3)\geq 0\Rightarrow x_3\geq-1
   \end{equation}
   Its non--negativity is also automatically fulfilled due to $\lVert\vec{x}\rVert=1$.
\end{proof}
\vspace*{.3em}
{\it Remark:} The condition $X\geq0$ is only in the $L=2$ case automatically fulfilled.  
\vspace*{.3em}

\noindent Next we will explicitely compute the matrix representation $(a_{ij})$ for the completely positive map $A(\cdot)$ 
in terms of the Pauli basis.\\
\begin{mylemma}
   The matrix representation of the completely positive map $A(\cdot)$ is the diagonal matrix
   \begin{equation*}
      \begin{split}
         (a_{ij})=\diag(\frac{1}{2},
         &(p_0+p_1)-\frac{1}{2},
         (p_0+p_2)-\frac{1}{2},\\
         &(p_0+p_3)-\frac{1}{2})
      \end{split}
   \end{equation*}
   \label{lemma:twobytwo:matrix}
\end{mylemma}

\begin{proof}
   The matrix elements $A(\cdot)$ with respect to the Pauli basis are given as
   \begin{equation}
      \begin{split}
         a_{kl}
         &=\frac{1}{4}\Trace{A(\sigma_k)\sigma_l}\\
         &=\frac{1}{4}\sum_np_n\Trace{\sigma_n\sigma_k\sigma_n^*\sigma_l}
         \defeq\frac{1}{4}\sum_np_n\Trace{\alpha_{nkl}}
      \end{split}
   \end{equation}
   Using the properties of the Pauli matrices one can compute
   \begin{equation}
      \begin{split}
         \Trace{\alpha_{nkl}}
         &=2\begin{cases}
            -\delta_{kl} & 0\neq n\neq l\neq0 \\
            +\delta_{kl} & \text{else}\\
         \end{cases}
      \end{split}
   \end{equation}
   which gives then
   \begin{equation}
      \begin{split}
         (a_{ij})=\frac{1}{2}\diag(1,\,
         &p_0+p_1-p_2-p_3,\\
         &p_0-p_1+p_2-p_3,\\
         &p_0-p_1-p_2+p_3)
      \end{split}
   \end{equation}
   Using now the normalization, i.e. $p_0=1-p_1-p_2-p_3$ gives the desired result.
\end{proof}
\vspace*{1em}
\begin{mytheorem}
   The solution of the problem in (\ref{eq:twobytwoproblem}) is
   \begin{equation}
      \begin{aligned}
         \max_{x,y\in\setblochrankone(2)}\langle x,ay\rangle=\frac{1}{2}(1+\max_{k=1,2,3}\{|2(p_0+p_k)-1|\})
      \end{aligned}
   \label{eq:twobytwosolution}
\end{equation}
\end{mytheorem}

\begin{proof}
   Using Lemma \ref{lemma:twobytwo:blochmanifold} and \ref{lemma:twobytwo:matrix} gives explicitely:
   \begin{equation}
      \begin{aligned}
         1\geq F&\defeq\max_{x,y\in\setblochrankone(2)}\langle x,ay\rangle\\
         &=\frac{1}{2}(x_0y_0+
         \max_{\lVert\vec{x}\rVert_2=\lVert\vec{y}\rVert_2=1}{\langle\vec{x},b\vec{y}\rangle})\\
         &=\frac{1}{2}(1+\max\{
         \underbrace{|2(p_0+p_1)-1|}_{x^\opt(1)=(1,1,0,0)},
         \underbrace{|2(p_0+p_2)-1|}_{x^\opt(2)=(1,0,1,0)},\\
         &\hspace*{6.5em}\underbrace{|2(p_0+p_3)-1|}_{x^\opt(3)=(1,0,0,1)}\})\geq\frac{1}{2}      \\
      \end{aligned}
      \label{eq:twobytwosolution}
   \end{equation}
   where $(b_{ij})$ is the lower right $3\times3$ sub--matrix of $a$. Because $(b_{ij})$ is a diagonal matrix,
   only the three optimal vectors  $x^\opt(n)$ given in (\ref{eq:twobytwosolution}) are possible. 
\end{proof}
\vspace*{1em}

Furthermore the $k$th component of the optimal equalizer is 
\begin{equation}
   y^\opt(n)_k=\sign\{2(p_0+p_n)-1\}x^\opt(n)_{k}
\end{equation}

\subsubsection{Discussion}
\noindent 
The simple solution (\ref{eq:twobytwosolution}) is well-suited to discuss several cases which occur
also in the general WSSUS pulse shaping problem.\\

\noindent(1) {\it non-dispersive channels}: 
If \mbox{$p_k=1$} for some $k=0\dots3$, this will yield
$F=1$ which is achieved with {\bf any} $x^\opt(n)$. This is the so called flat fading 
(non-selective) channel. The maximal channel fidelity (''$F=1$'') can be achieved. No precoding is needed. \\

\noindent(2) {\it single-dispersive channels}: If \mbox{$p_k=p_l=0$} for some $k,l=0\dots3$ and $k\neq l$
yields again $F=1$ which achieved with $x^\opt(n)$ where $n\neq k$ and $n\neq l$. This is for example
for $k=2$ and $l=3$ the frequency selective, time--invariant channel. All contributions $\Shift_\mu$ can 
be diagonalized simultaneously which is achieved for example in OFDM. 
The maximal channel fidelity (''$F=1$'') is achieved again (in practice there is 
still a fidelity loss for OFDM due to the cyclic prefix).\\

\noindent(3) {\it doubly--dispersive ''underspread'' channels}: If in general only 
\mbox{$p_k>\frac{1}{2}$} for some $k=0\dots3$ we have only
$F>\frac{1}{2}$.  But if $k=0$ follows 
$F=p_0+p_n$ where $n=\arg\max_{m=1,2,3}\{p_m\}$, achieved with $x^\opt(n)$.
A closer inspection shows, that if $k\neq0$ the solution will depend on some ordering
property of the scattering powers. The optimal precoder depends 
now explicity on multiple values $p_m$.\\

\noindent(4) {\it Completely overspread channels}: If \mbox{$p_k=\frac{1}{4}$} for all $k$
the channel fidelity gives $F=\frac{1}{2}$ which is achieved with {\bf any} 
$x^\opt(n)$. This is the worst case scenario. \\

\noindent It is quite interesting what happens if we fix the scattering power $p_0$, i.e.
to consider $F$ as the function $F(p_0,\vec{p})$ where $\vec{p}=(p_1,p_2,p_3)$
\begin{equation}
   F(p_0,\vec{p})=\frac{1}{2}(1+\max_{k=1,2,3}\{|2(p_0-p_k)-1|\})
\end{equation}
Clearly, $F(p_0,\vec{p})$ is jointly and separately convex in $p_0$ and $\vec{p}$.
Furthermore  is $F(p_0,\vec{p})=F(p_0,\Pi\vec{p})$ for every permutation $\Pi$, so $F$ is Schur-convex
in the second argument (see for example \cite{bhatia:matrixanalysis}), i.e. for a fixed $p_0$ follows
\begin{equation}
   \begin{split}
      \vec{p}_1\major\vec{p}_2\Rightarrow F(p_0,\vec{p}_1)\geq F(p_0,\vec{p}_2)
   \end{split}
\end{equation}
Using Schur-convexity for every fixed $p_0$ follows:\\
\noindent(5) {\it the worst case channel}: is given for
$p_k=\frac{1-p_0}{3}$ for all $k=1,2,3$ yielding 
\begin{equation}
   \begin{split}
      &\min_{\lVert\vec{p}\rVert_1=1-p_0} F(p_0,\vec{p})
      =F(p_0,\frac{1-p_0}{3}(1,1,1))\\
      &=\frac{1}{2}(1+\frac{1}{3}|4p_0-1|)
      =\frac{1}{2}+\frac{2}{3}|p_0-\frac{1}{4}|
   \end{split}
\end{equation}
Note that in the quantum context this corresponds to the {\it general
depolarizing channel} or {\it Lie algebra channel} \cite{arxiv:0502153}.\\
\noindent(6) {\it the best case channel}: is given for
$\vec{p_k}=(1-p_0)\vec{e}_k$ where $\vec{e}_k$ is a standard basis vector
and $k\in\{1,2,3\}$ yielding 
\begin{equation}
   \begin{split}
      \max_{\lVert\vec{p}\rVert_1=1-p_0} &F(p_0,\vec{p})
      =F(p_0,(1-p_0)\vec{e}_k)\\
      &=\frac{1}{2}(1+\max\{|2p_0-1|,1\})=1
   \end{split}
\end{equation}
achieved with $x^\opt(k)$. This is again an single-dispersive channel
because $\sigma_0$ commutes with $\sigma_k$.

\subsubsection{Construction of the precoders}   
In this part we will explicitely calculate the corresponding matrix representation
$X^\opt(n)$ from the Bloch parameterizations $x^\opt(n)$ 
using $X=\frac{1}{2}\sum_i x_i\sigma_i$ which gives 
\begin{equation}
   \begin{split}
      X^\opt(n)&=\frac{1}{2}(\sigma_0+\sigma_n)\\
      Y^\opt(n)&=\frac{1}{2}(\sigma_0\pm\sigma_n)
   \end{split}
\end{equation}
These matrices are the rank-one projectors onto the optimal precoders in the
original problem. Hence, turning back to
precoders and equalizers $x(n),y(n)\in\mathbb{C}^2$ it can be verified that
($X^\opt(n)$ similarly)
\begin{equation}
   \begin{aligned}
      Y^\opt(1)&
      =\frac{1}{\sqrt{2}}{\small \left(\begin{array}{c}           
             \pm1 \\
             1 \\
          \end{array}\right)}\frac{1}{\sqrt{2}}{\small\left(\begin{array}{c}           
             \pm1 \\
             1 \\
          \end{array}\right)}^*\\      
      Y^\opt(2)&
      =\frac{1}{\sqrt{2}}{\small \left(\begin{array}{c}           
             \pm 1 \\
             i \\
          \end{array}\right)}\frac{1}{\sqrt{2}}{\small\left(\begin{array}{c}           
             \pm 1 \\
             i \\
          \end{array}\right)}^*\\
      Y^\opt(3)&
      =\frac{1}{\sqrt{2}}{\small\left(\begin{array}{c}           
             1/2(1\mp1) \\
             1/2(1\pm1) \\
          \end{array}\right)}\frac{1}{\sqrt{2}}{\small\left(\begin{array}{c}           
             1/2(1\mp1) \\
             1/2(1\pm1) \\
          \end{array}\right)}^*
   \end{aligned}
\end{equation}
In the following I will use w.l.o.g. the $\pm$-version, hence $X^\opt(n)=Y^\opt(n)$. 
The following precoders are then solutions of the optimization problem
\begin{equation}
   \begin{aligned}
      x(1)={\frac{1}{\sqrt{2}}}\left(\begin{array}{c}           
           1 \\
           1 \\
        \end{array}\right) \quad& 
      Fx(1)=\left(\begin{array}{c}           
           1 \\
           0 \\
        \end{array}\right) \\
      x(2)={\frac{1}{\sqrt{2}}}\left(\begin{array}{c}           
           1 \\
           i \\
        \end{array}\right) \quad& 
      Fx(2)={\frac{1}{2}}\left(\begin{array}{c}           
           1+i \\
           1-i \\
        \end{array}\right) \\
      x(3)=\left(\begin{array}{c}           
           0 \\
           1 \\
        \end{array}\right) \quad&
      Fx(3)={\frac{1}{\sqrt{2}}}\left(\begin{array}{c}           
           1 \\
           -1 \\
        \end{array}\right)
   \end{aligned}
\end{equation}
The solution $x(1)$ is maximally localized in the frequency domain, i.e. completely spread
out in the time domain. The reverse case holds for $x(3)$. To understand $x(2)$ one has 
to perform a rotation in the time-frequency plane.

\subsection{The Multiplexing Scheme}
Solving the channel fidelity problem as done in the previous section is the first step
toward optimal signatures for a given WSSUS channel statistics. In this part we will discuss
now how multiplexing has be performed. In $\mathbb{C}^2$--case at spectral efficiency 
\mbox{$\det\Lambda^{-1}=1$} this means multiplexing of a second data stream only. Recall that
with Gabor (Weyl--Heisenberg) signaling the same channel fidelity
is achieved for the second data stream. It remains first to select the lattice with minimum
averaged interference and then perform some kind of orthogonalization (''tighten'') 
procedure.  But the $\mathbb{C}^2$ case admits already the following transmitter side orthogonality relations
\begin{equation}
   \begin{split}
      \langle x(n),\Shift_{(0,1)}x(n)\rangle
      &=\frac{1}{2}\Trace{\sigma_3(\sigma_0+\sigma_n)}=\delta_{n3}\\
      \langle x(n),\Shift_{(1,0)}x(n)\rangle
      &=\frac{1}{2}\Trace{\sigma_1(\sigma_0+\sigma_n)}=\delta_{n1}\\
      \langle x(n),\Shift_{(1,1)}x(n)\rangle
      &=\frac{i}{2}\Trace{\sigma_2(\sigma_0+\sigma_n)}=i\delta_{n2}\\
   \end{split}
\end{equation}
Hence, for each channel optimal pulse ''$n$'' there are several schemes ''$n$'' 
for multiplexing a second data stream which admits transmitter--side orthogonality, i.e. no further 
orthogonalization is required. The ''channel fidelity''--optimal precoder is also $\SINR$--optimal.
If for example the channel optimal precoder is $x(3)$,  ''time-division multiplexing'' 
via $\Shift_{(0,1)}$ is one of the optimal schemes. For $x(1)$ in turn ''frequency-division multiplexing''
is the right scheme.

\section{Conclusions}
In this article new insight into the \mbox{WSSUS}
pulse shaping problem are given from a precoding viewpoint. 
The precoding problem is solved for a very simple class of random $2\times2$ channels under the
assumption that the transmitter
has only knowledge of the second order statistics and the receiver has full knowledge of the channel.
It is observed that optimality in the channel--fidelity sense is achieved with concentration  in a certain domain
similarly as one would expect from the continuous case. Unfortunately the direct extension of 
this approach to higher dimension is problematic due to the difficulties arising with the Bloch parameterization.
But a staggered extension could be conceivably and will probably studied in future.

\bibliographystyle{IEEEtran}
\bibliography{references}

\end{document}